# An educational distributed Cosmic Ray detector network based on ArduSiPM.

**V Bocci, G Chiodi, P Fresch , F Iacoangeli, L Recchia,**

INFN sezione di Roma, P.le Aldo Moro 2, Roma Italy

E-mail:Valerio.Bocci@roma1.infn.it

**Abstract.** The advent of microcontrollers with enough CPU power and with analog and digital peripherals makes possible to design a complete particle detector with relative acquisition system around one microcontroller chip. The existence of a world wide data infrastructure as internet allows for devising a distributed network of cheap detectors capable to elaborate and send data or respond to settings commands. The internet infrastructure enables to distribute the absolute time (with precision of few milliseconds), to the simple devices far apart, with few milliseconds precision, from a few meters to thousands of kilometres. So it is possible to create a crowdsourcing experiment of citizen science that use small scintillation-based particle detectors to monitor the high energetic cosmic ray and the radiation environment.

## 1. Introduction

The outreach and education programs can greatly benefit of the incredible advances in technology. Regarding the photon detection, in the last years, a very small and relatively cheap solid devices, are available: the SiPM (Silicon Photon Multiplier)[1]. Also known as MPPC (multi pixel photon counter) [2], it is a solid photon counting with single photon detection skill. The SiPM/MPPC is made up of an avalanche photodiode (APD) array on common Si substrate, each single element working in Geiger mode. When a photon hits the silicon, it produces a photoelectron through photoelectric effect in one of the microcells. Respect to the classical Photo Multiplier, the SiPM is more compact, more robust and does not necessitate of in the KV order bias voltage that is difficult to manage in compact design. The necessity to use expensive and fragile Photo Multipliers (PM) as photon detector limited, in the past, the use of scintillation materials in radiation detector handled device preferring the Geiger tube approach.
The first modern scintillator detector for alpha particles was built using a Zinc sulphide screen, as scintillator, coupled with a photo multiplier by Curran and Baker in 1944. Their work was described in classified report in 1944, disclosed in 1948 [3].
In our time, the scintillation counter are used extensively, in nuclear and particle physics, to build up detectors as fast trigger, calorimeters, tracker, Time of flight. The idea to couple a scintillator with a modern SiPM start in the early 2000s first application , in small animal positron emission tomography, comes up in 2006[4]. The acquisition of a single Silicon Photomultiplier can require electronics modules as: preamplifier, discriminator, bias voltage power supply, temperature monitor, Scalers, Analog to Digital Converter and Time to Digital Converter. Since 2011, our group start to build a conceptually simple system to acquire and control SiPM photodetectors mainly to create an handy particle detector. The idea was to utilize one of the best off-the-shelf microcontrollers and to utilize the CPU and all the peripheral inside the chip with a minimal signal conditioning outside. We realized a system based on the Arduino DUE [5] development board and a specific shield: the ArduSiPM[6]. The first target of this particle detector was as scientific instrumentation for health probe [7] and a trigger for the CERN experiment UA9 [8]. The use of Arduino DUE, a well know platform in the educational environment, and the potentially low cost of the system convince us to transfer the design for an outreach and educational use.

In this paper we show the idea to add to ArduSiPM a very low cost external network processor to communicate using IoT technology like MQTT (MQ Telemetry Transport) protocol and NTP (Network Time Protocol) building up a crowdsourcing experiment of citizen science.

## 2. The ArduSiPM Hardware

The ArduSiPM consist of an Arduino Due Board, an ArduSiPM Shield, a Silicon Photomultiplier, a scintillator and an external Wi-Fi interface (Fig. 2). Arduino DUE is an open hardware and software development board based on the Atmel SAM3X8E ARM Cortex-M3 CPU. It is an off-the-shelf board widespread in the makers and education community with a free integrated development environment

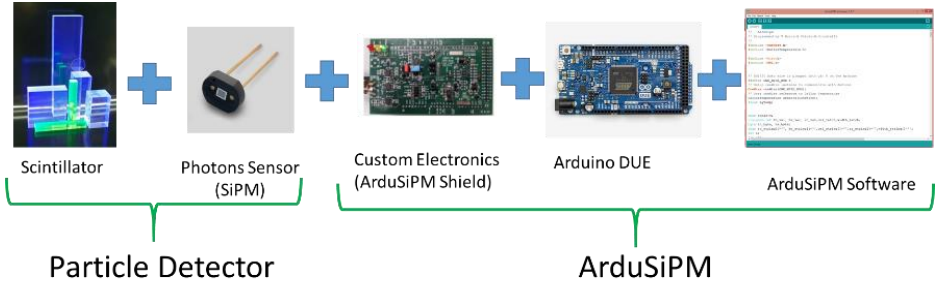

*Fig. 2 Main components of the ArduSiPM detector.*

(IDE). It is used widespread for didactics platform due to an easy to use program language. All the information about the platform are open and the schematics and the development software is available over the internet. The ArduSiPM Shield is our custom designed board with all electronics interface from Arduino DUE and a SiPM photodetector. The ArduSiPM Shield plugged in the Arduino DUE create an easily transportable device including Front-end electronics and data acquisition system. The global architecture of the system is in (Fig. 1). The SiPM and the temperature sensor are externally connected whereas the digital controlled bias supply, the voltage amplifier, a fast discriminator with programmable threshold, a peak-hold circuit, leds for monitoring and all outputs from analog circuit and digital controls are connected to the Arduino DUE.

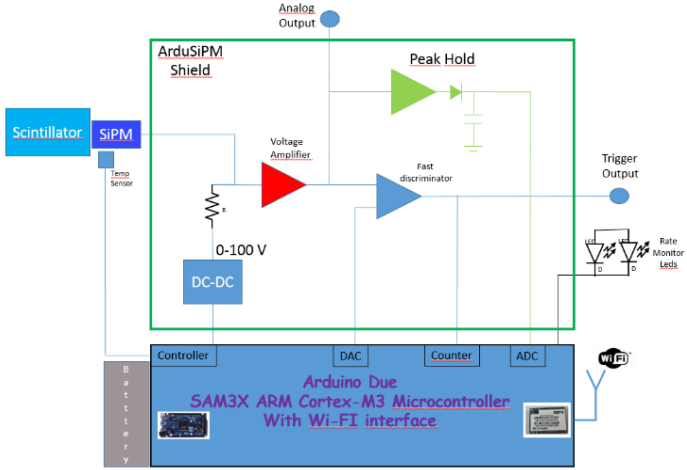

*Fig. 1 ArduSiPM global Architecture.*

The main components of the ArduSiPM Shield are:

*2.2. SiPM Power Supply adjustment circuit.*

A DC-DC converter use the Arduino 5 V to supply bias voltage in the range of 30-100 V. Once we fixed the nominal voltage it is possible to vary around this value of few Volt with 8 bits resolution. In combination with the temperature sensor we can real-time compensate the gain variation typical of this kind of detector.

*2.3. Analog circuit:Voltage Amplifier and Fast Discriminator.*

The voltage amplifier operate a signal conditioning of the SiPM adapting the output to the fast discriminator and the Arduino Analog to Digital converter range. The Amplifier is fast and linear enough to follow the SiPM response of few ns and to cover all the range of SiPM signal. The noise is less than a single SiPM one pixel signal. For monitoring purpose a replicated output of the amplifier is available as external analog connection. The SiPM's output Signal is very short (few ns). A fast 7 ns discriminator is used to discriminate the over threshold signals and to count them using Arduino DUE counter. The threshold is digitally controlled and his value is monitored using the internal ADC. The width of discriminator output is programmable to avoid after pulse counting and to control the death time of the pulse acquisition. A replicated TTL output of the fast discriminator is available as Trigger Output, so it can be used as trigger for external acquisition system.

*2.4. Peak height measurement.*

A precise circuit with fast peak detector is used as peak-hold. A sampling comparator features a very short switching charge to measure the short pulses coming out from the SiPM detector when the pulse generate a trigger. The Pulses are stretched over 1 μs to be converted from the 1 MSPS 12 Bits ADC of the SAM3X8E. A programmable digital signal control the fast discharge circuit to reset the circuit of the peak hold and to rearm the system for a new acquisition

### 3. The ArduSiPM network interface

The ArduSiPM Shield sends data over serial RS232, in the first prototype we choose a simple RS232-WI-FI interface to transmit data over the network. In the meantime the availability of powerful and cheap microcontrollers with Wi-Fi interface give to us the possibility to expand the network feature of ArduSiPM (Fig. 3). We attach a fully programmable network processor to handle all the communications tasks and external sensors without charge the main processor board. We chose the Espressif ESP8266 chip available at low price and with very powerful specification [9]. The Espressif ESP8266 is a System-On-A-Chip of few square millimetre with a 32-bit RISC CPU Tensilica Xtensa LX106 running at 80 MHz, with 64 KB of instruction RAM , with96 KB of data RAM and with QSPI flash supporting upto 16 MB. There is inside a complete IEEE 802.11 b/g/n Wi-Fi interface with WEP or WPA/WPA2 authentication, or open networks. There are 16 GPIO pins, SPI and I2C interface and a UART for RS232 serial communication. The chip comes out in 2014 with a Software development Kit and, from April 2015, there was a free porting of the firmware to run Arduino code. The availability of the Arduino IDE for ESP8266 gives to us the possibility to benefit of public domain code developed for Arduino, especially regarding the Internet Of Things (IoT) the Network Time Protocol (NTP) and webserver.

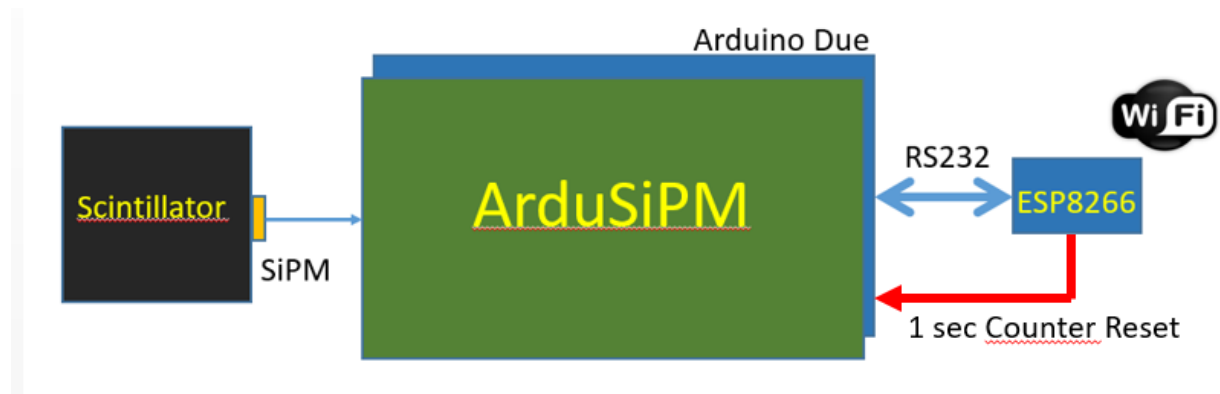

Fig. 3 The ArduSiPM detector connection to ESP8266 network processor. The RS232 send data and receive settings to the ESP8266. During time synchronization the ESP8266 send a reset signal to synchronize one sec ArduSiPM internal clock with NTP time.

*3.1. Use of MQTT protocol with ArduSIPM ESP8266 module.*

MQTT (Message Queue Telemetry Transport) is "lightweight" messaging protocol for use on top of the TCP/IP protocol. MQTT was invented in 1999 by Andy Stanford-Clark (IBM) and Arlen Nipper (Arcom, now Cirrus Link), their goal was to create a protocol for minimal battery loss and minimal bandwidth connecting oil pipelines over satellite connection to interface sensors to a SCADA system. The protocol come a ISO/IEC standard in 2016 (ISO/IEC 20922:2016)[10]. It uses a publish/subscribe architecture in contrast to HTTP with its request/response paradigm. Publish/Subscribe is event-driven and enables messages to be pushed to clients. The MQTT is light messaging protocol but also light from the processor time required. There are different version of MQTT library for the Arduino ESP8266, at the moment there is not a full protocol library. We choice the Adafruit MQTT client library that cover the main parts of the protocol. Every second the ArduSiPM provide the number of events and the time in microseconds of each event relative to a second lap (20 bits). The ESP8266 build up a MQTT event packet adding: the serial number of ArduSiPM (2 byte), the NTP seconds from the internal real time clock (4 bytes), the 4 bytes fraction of second are built using as MSB the 20 bits time from ArduSiPM the 12 LSB bits are forced to zero but are available in case of a better time reference like GPS (Fig. 4), The MQTT rate data packet is built with 2 bytes for ArduSiPM serial number and 2 bytes for data rate. Other MQTT messages are built using sensors connected to ESP 8266 (temperature, humidity, pressure…) and transmitted at lower rates.

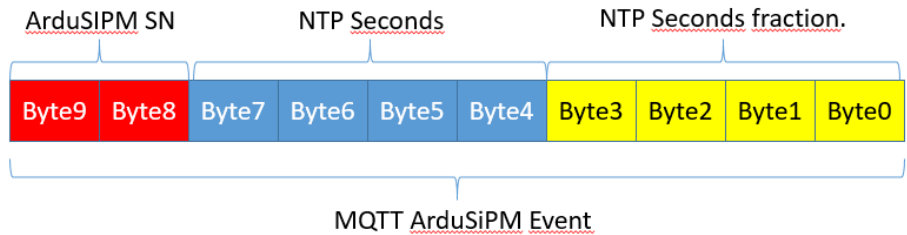

*Fig. 4:ArduSiPM Event packet transmitted via MQTT publish to MQTT broker. Each packet report ArduSiPM Serial Number and the NTP time of the Event. The second fraction is computed inside ArduSIPM the NTP time in seconds come from the ESP8266 internal clock.*

### 3.2. NTP protocol with ArduSIPM ESP8266 module.

Network Time Protocol (NTP) is a networking protocol, for clock synchronization between computer systems, over packet-switched, variable-latency data networks. NTP was designed by David L. Mills of the University of Delaware [11]. NTP is intended to synchronize all participating computers to within a few milliseconds of Coordinated Universal Time (UTC). A simplified version of the NTP protocol is SNTP easy to implement in the ESP8266 processor. Using SNTP the ArduSiPM ESP8266 module receive from internet the time with a few milliseconds precision, in correspondence of the lap second the ESP8266 set the time and send a hardware counter reset to ArduSiPM. In this way the one second counter inside the ArduSiPM is synchronized with the NTP time (*Fig. 3*).
The NTP system is economic and it come for free using the ESP8266, but better time synchronization at nanoseconds level is possible using a Global Position System (GPS) receiver with one Pulse Per Second (1 PPS) signal, the use of GPS presupposes the sky view availability or a long connection between ArduSiPM and GPS module.

### 3.3. WEB server and configuration interface

The four task managed from ArduSiPM ESP8266 module are: Wi-Fi connection and communication, Management of NTP time and ArduSiPM clock synchronization, MQTT client data formatting and transmission, readout of environment sensors. A simple web interface run in the module for setting variable parameters like: Wi-FI hotspot, MQTT Broker, NTP Server.

## 4. The MQTT broker.

The central communication point in the MQTT protocol is the broker (the server in MQTT), it is in charge of handle publish/subscribe protocol. A broker can handle up to thousands of concurrently connected MQTT clients. The broker is primarily responsible for receiving all messages, filtering them, decide who is interested in it and then sending the message to all subscribed clients.

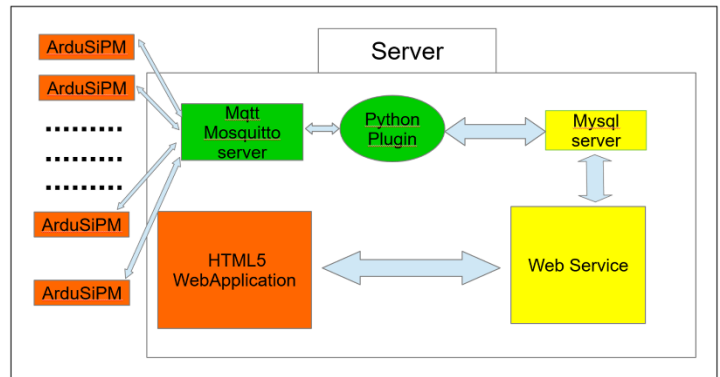

*Fig. 5 MQTT broker connected to MySQL server for data storage. The data can be accessed from HTM5 web application using a Web Service.*

We choose Mosquitto [12] as MQTT broker installed in a linux server. Mosquitto is an open source (EPL/EDL licensed) message broker that implements the MQTT. Each ArduSiPM, using his username and password, send the data to the MQTT Broker by means of a Python plugin program communicate and record any single data from any ArduSiPM inside a MySql database. A web service interfaced to the database can access the data and can publish using a html5 web application (Fig. 5). The idea is to provide a subset of data or the all dataset to the ArduSiPM community for visualization and elaboration.

### 5. A Cosmic ray and environment radiation crowdsourcing experiment of citizen science

The first observation of a cosmic ray particle with an energy exceeding $1.0\times10^{20}$ eV (16 J) made by Dr John D Linsley and Livio Scarsi at the Volcano Ranch experiment in New Mexico in 1962 has opened a new frontier in the research of Ultra High Energetic Cosmic Rays (UHECR)[13] . The size of the shower on the ground is linked to energy of the primary cosmic ray this implies the construction of detectors extended over a large area to sample the cosmic shower. With the infrastructure above described it is possible to create a "citizen science" experiment in a large area. Each ArduSiPM publish online the number of event for each seconds and the absolute time with a precision of few millisecond for each event in case of NTP timing or microsecond in case of GPS timing (Fig. 6). This infrastructure does not claim to be competitive with modern large area cosmic ray experiment like Auger (Argentina), Telescope Array (Utah) but a project to offers students the opportunity to understand how detectors, data acquisition, Ultra Energetic cosmic ray research work.  The detector can be also used as environment monitor of radiation and changing the scintillator type it is possible to be sensible to particular type of radiation as described in [14].

### 6. Conclusions

We describe the use of the small scintillation detector ArduSiPM to build up a citizen science experiment for Ultra High Energetic Cosmic Rays and radiation environment monitor. All the architecture of the experiment is described involving data acquisition, timing distribution data storage using as connection of different devices the internet availability. The ArduSiPM device from 2016 is a commercially available from an Italian company [15] under INFN license, this can potentially create a widespread use of the device especially in schools, universities, amateur or nonprofessional scientists.

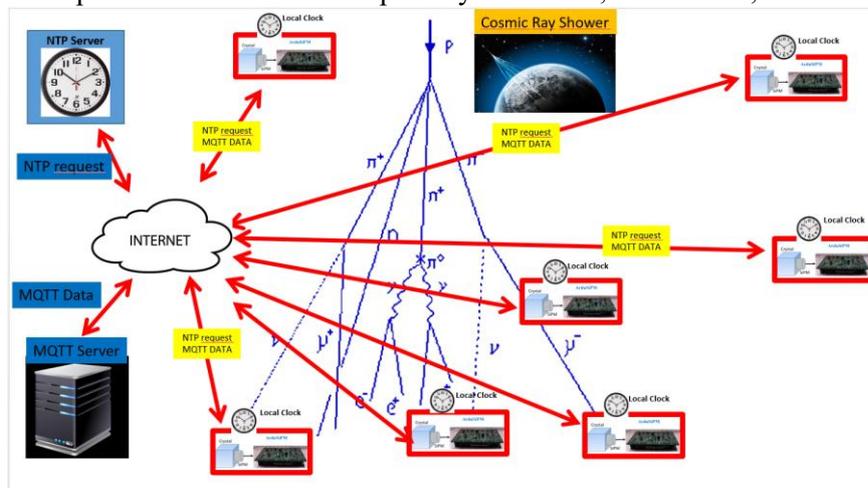

*Fig. 6 Scheme of the ArduSiPM crowdsourcing citizen science experiment for the research of Ultra High Energetic Cosmic Rays and radiation monitor*

### 7. Acknowledgement

We want to thanks Matteo Bocci for the proof of concept of the MQTT to MySQL data bridge, Cino Matacotta , Ilaria Giammarioli, Pier Paolo Deminicis of INFN technology transfer, Catia Peduto, Francesca Scianniti, Barbara Sciascia of the INFN Magazine Asimmetrie and INFN communication office, Roberta Santacesaria of LHCb group coordinator of INFN Rome and Antonella and Incicchitti Astro Particle Physics coordinator of INFN Rome, Armando Paliani of Robot Domestici for having believed and supported the project.